# Transformable Soft Quantum Device based on Liquid Metals with Sandwiched Liquid Junctions

Xi Zhao,[1,2] Jianbo Tang,[3] Yongze Yu,[1] and Jing Liu,[1,2,3,4]*

[1]Technical Institute of Physics and Chemistry, Chinese Academy of Sciences, Beijing 100190, China.

[2]School of Future Technology, University of Chinese Academy of Sciences, Beijing 100049, China.

[3]Department of Biomedical Engineering, School of Medicine, Tsinghua University, Beijing 100084, China.

[4]Institute for Advanced Study on Liquid Metal, Yunnan University, Yunnan Province 650504, China

*Corresponding author. Email: jliu@mail.ipc.ac.cn.

**Abstract:** Quantum tunneling effect has been an important issue in both fundamental science and practical applications. Classical quantum tunneling devices are generally rigid in structures which may encounter technical difficulties during fabrication, functional tuning and shape adapting. Here through introducing the room-temperature liquid metals as two conductive electrodes and a soft even liquid insulating layer sandwiched between them, we proposed a conceptually innovative all-soft or liquid quantum device which would help realize a couple of unconventional quantum capabilities such as flexibility, deformability, transformability, and reconfigurability, etc. which may not be easily offered by a rigid quantum device. Representative structural configurations to make such transformable quantum devices via sandwiching various liquid metal-insulating layer-liquid metal (LM-IL-LM) components are suggested. The feasibility for making such an all-soft quantum device is interpreted through experimental evidences and theoretical evaluations. Potential future applications of the proposed devices in a group of emerging fields including intelligent quantum systems and quantum computing are prospected.

## 1. Introduction

It is a well-known fact that if there exists an insulating layer between two conductors (or semiconductors and superconductors), electrons are not able to pass through the layer from one side to the other. The insulating layer here serves as a barrier, or called potential junction, to electrons. However, when the thickness of such insulating layer becomes comparable to the de Broglie wavelength, electrons can pass through the barrier [1]. This special phenomenon is called quantum tunneling effect. In fact, the quantum tunneling effect plays an essential role in numerous fundamental physical phenomena such as the nuclear fusion taking place in the sun [2] and the biological photosynthesis process [3]. Moreover, quantum tunneling theory is also widely used in many fields [4, 5] such as scanning tunneling microscopy, optics, semiconductor physics and superconductor physics (Fig. 1) etc.

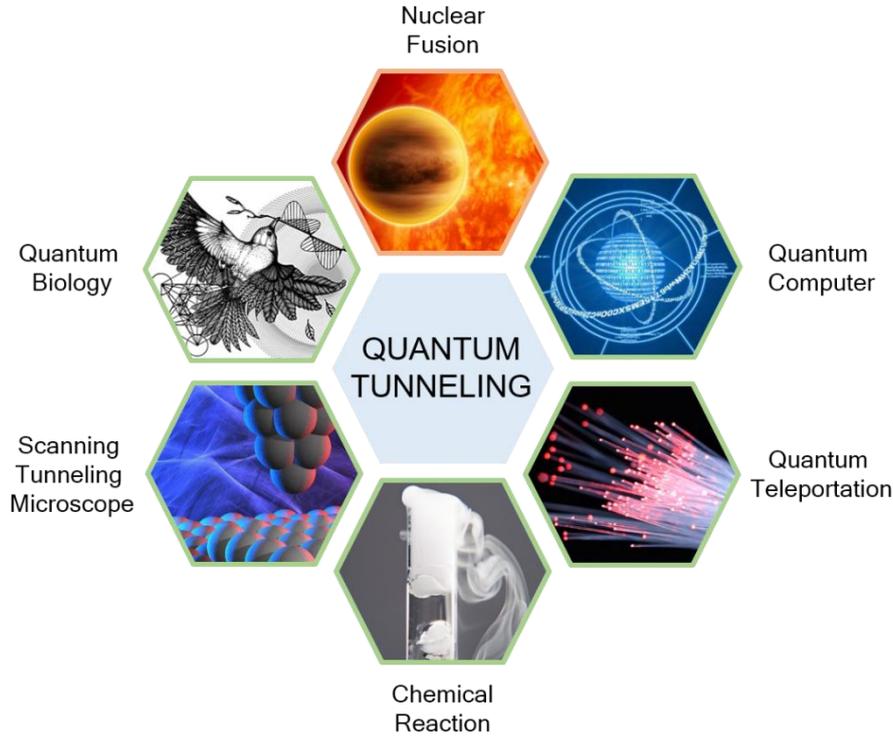

**Fig. 1.** Typical situations where quantum tunneling effect plays important role.

So far, nearly all the quantum tunneling devices are composed of a rigid sandwich structure with a thin insulating layer (0.1~10 nm thickness) inserted between two electrodes on its both sides [6]. Since these components are all solid, the device structure will not be readily adjusted. Thus the entire device cannot be reshaped, split or reassembled either. Once fabricated, such devices would only be able to perform certain specified functions restricted by its predefined structure. This gives rise to a basic question that could such device be transformable in shape and also tunable in its quantum functions? Clearly, if the building components of a quantum tunneling device can all be replaced by specific liquid or soft materials, all-soft even liquid quantum tunneling device could be realized, which may provide totally different performances compared to the traditional rigid ones. Such soft feature would bring about possibilities for quantum tunneling devices to realize novel capabilities such as flexibility, deformability, transformability, and controllability.

As the initial trial, a class of distinguishing candidates for the conductive soft electrode material could be room-temperature liquid metals. Among them, the gallium-based liquid metals (LM) especially combine the virtues of liquid (fluidity, flexibility, and deformability) and the advantages of metals (high electrical and heat conductivity) [7]. Additionally, it has been disclosed recently that the LMs would display a series of intriguing interfacial phenomena in different types of medium, such as acid solution [8, 9], basic solution [10-13], surfactant solution [14], and organic gels [15], etc. When immersed in a specific solution, a LM-IL-LM sandwich structure forms spontaneously, which reminds us that such structure would lead to a potential quantum tunneling device given that the insulating layer could reach the characteristic length ($L_C$) for the quantum tunneling effect. It is based on this basic consideration, we are dedicated here to propose a new conceptual quantum device which can be termed as all-soft or all liquid quantum device (A-LQD) to innovate the conventional rigid ones. The highly transformable performance of such kind of quantum device is more conducive to the popularization and application of future low-cost and intelligent quantum technology.

## 2. Quantum tunneling principles

The quantum tunneling effect can be explained in terms of uncertainty relations [16, 17] and the wave–particle duality [18, 19]. Considering the case of a quantum tunneling junction with a rectangular barrier (Fig. 2), a particle (e.g., electron) with energy $E(> 0)$ is shot toward the barrier in the x-axis direction. The potential function of the barrier is:

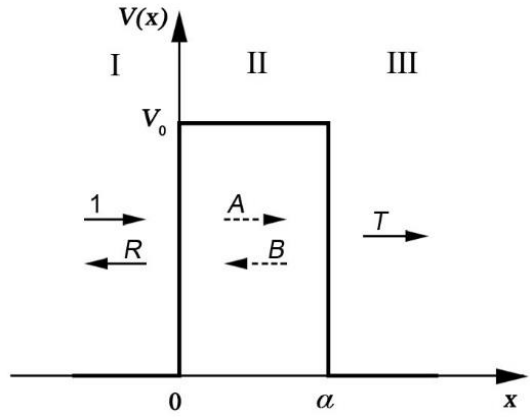

**Fig. 2.** The schematic of a quantum tunneling junction with a rectangular barrier.

$$V(\text{x}) = \begin{cases} 0, & x < 0 \quad (\text{I}) \\ V_0, & 0 < x < a \quad (\text{II}) \\ 0, & x > a \quad (\text{III}) \end{cases} \quad (1)$$

Using the treatment of wave–particle duality of matter in quantum mechanics, the Schrödinger equation for a steady state implies [5]:

$$\frac{\hbar^2}{2m}\frac{d^2\Psi(x)}{dx^2} = [V(x) - E]\Psi(x) \quad (2)$$

Assuming the wave is incident from the left zone I ($x = -\infty$), a portion of the wave will be reflected at the left side of the barrier ($x = 0$) while the rest enters the barrier (zone II). Likewise, a similar process takes place at the right hand side of zone II ($x = a$). As a result, a portion of wave overcomes the barrier and enters zone III. The wave functions in the three zones can be expressed as follows:

$$\begin{cases} \Psi_\text{I}(x) = e^{ipx/\hbar} + Re^{-ipx/\hbar} \\ \Psi_\text{II}(x) = Ae^{ip'x/\hbar} + Be^{-ip'x/\hbar} \\ \Psi_\text{III}(x) = Te^{ipx/\hbar} \end{cases} \quad (3)$$

where $p = \sqrt{2mE}$, $p' = \sqrt{2m(E - V_0)}$, $R, T, A, B$ are parameters determined by the convergence condition of the wave function on both sides of the barrier ($x = 0, x =$ $a$). Thus the reflection coefficient and the transmission coefficient of the barrier are $|R^2|$ and $|T^2|$, respectively. Since $\Psi(x)$ and $\Psi'(x)$ should be continuous at $x = 0$ and $x = a$, one has:

$$|T^2| \approx \frac{16E(V_0-E)}{V_0^2} e^{-2\beta a} \quad (4)$$

where $\beta = \frac{|p'|}{\hbar} = \frac{\sqrt{2m(V_0-E)}}{\hbar}$.

It can be seen from Eq. (4) that the particle has certain possibility to pass through the barrier, and the tunneling probability decreases exponentially with the increase of the barrier height $a$. And if $a$ could be adjusted on demand, the tunneling possibility of particles at the junction could also be controlled, resulting in intelligently tunable performance of the quantum tunneling devices.

## 3. Configurations of quantum device

### 3.1 Rigid and soft quantum device

As indicated in Fig. 3(a), the sandwich structure of a traditional rigid quantum tunneling device is typically composed of three parts: two solid electrodes and an insulating layer. The intermediate barrier may consist of a non-superconducting metal, a thin layer of insulating liquid or air. Since tunneling effect occurs with barriers of thickness about 0.1~10 nm, it usually requires precise manufacturing of the electrodes. Besides, solid surfaces suffer from wear and corrosion, thereby reducing the reliability and durability of the devices.

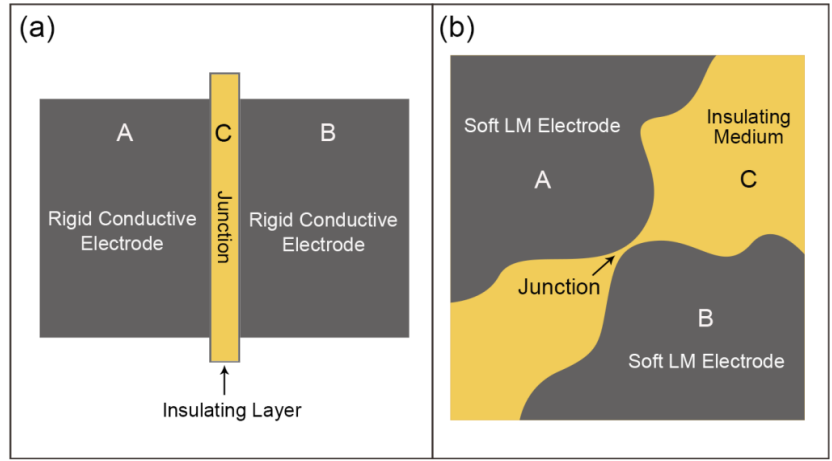

**Fig. 3.** The schematic of (a) rigid quantum device and (b) all-soft quantum device.

As explained in the former section, all-soft quantum devices can be obtained when replacing solid conductors with conducting liquid electrodes made of room-temperature gallium-based LMs (Fig. 3(b)). Without losing generality, such insulating layer can be water, oils, surfactant solutions, hydrogels and many other soft materials. Distinct from the traditional rigid-body quantum devices which cannot change their structures to implement different applications, the tunneling junction of the present all-soft quantum device can be easily transformed and manipulated through applying external forces or fields, thus exhibiting high flexibility and controllability. Soft quantum tunneling devices will not be limited by the traditional sandwich structure and they can also work on various configurations.

### 3.2 All-liquid quantum device (A-LQD) with an insulating liquid layer

If appropriate components are adopted, an all-liquid quantum device (A-LQD) can be constructed. In a recent study of a LM-electrolyte system, we have found that LM droplets could persistently float on the same LM bath without coalescence when subject to an electric field [20]. The electrolyte here serves as a solvent to remove the oxide layer on the LM so as to maintain a perfect smooth surface. A thin running liquid layer preventing the droplet from direct contacting with the lower LM bath was found to be responsible for such phenomenon (Fig. 4(a)). A stable thickness of the liquid layer $e_0$ was determined to be of micrometer scale and it was a result of the balance between the lubricating force ($F_L$) and gravity ($G$). For a transient case, a nanometer or even infinitely small scale gap can easily be realized over the coalesce process between the upper metal droplet and the lower liquid metal pool. According to the lubricating theory [21], the lubricating force of the liquid layer reads as

$$F_L \propto \mu_E v R^3 / e_0^2 \quad (5)$$

where $\mu_E$ is the dynamic viscosity of the liquid, $v$ is its velocity, and $R$ is the droplet radius. As can be seen from Eq. (5), $e_0$ is a highly adjustable parameter and it can be minimized by using smaller droplets or controlling other parameters, though its usual length scale is beyond that of $L_C$.

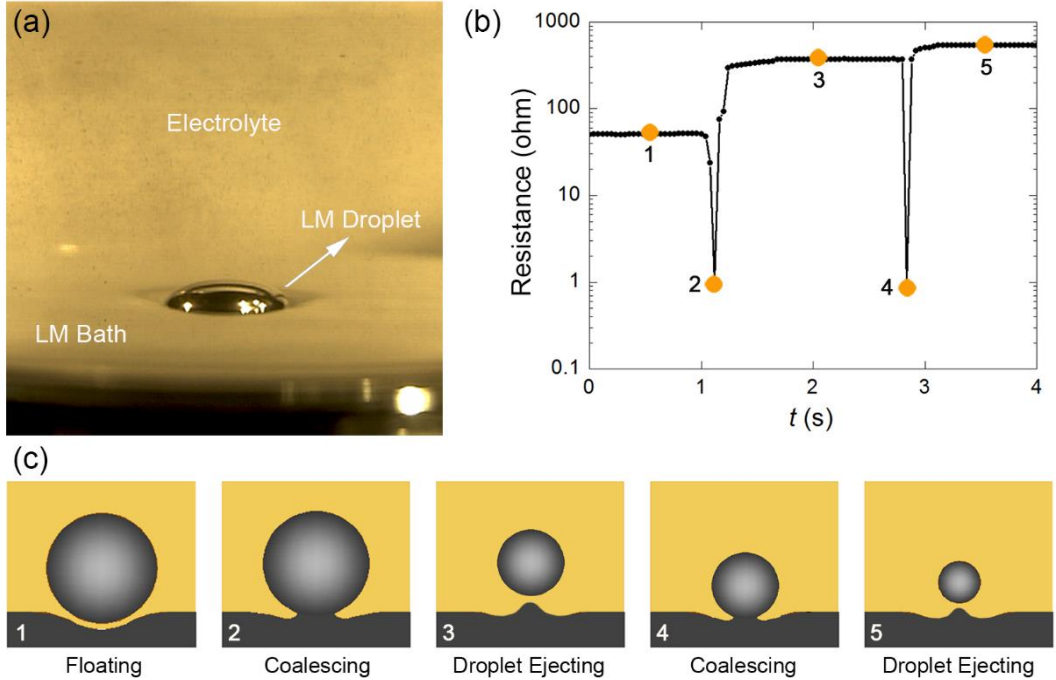

**Fig. 4.** (a) A LM droplet floating on a LM bath separated by a liquid layer (cannot be seen from the image) under an external electrical field; (b) Resistance evolution of the liquid layer after cutting off the applied voltage; (c) Schematics of the coalescence of a LM droplet and the successive satellite droplet ejecting with corresponding stages indicated in Fig. 4(b). Note that original data comes from the experiments in [20].

Clearly, once the electric field was cut down, the levitating LM droplet became unstable and merged with the LM bath. The droplet then coalesced with the LM bath accompanied with satellite droplet ejecting. As can be seen from our experimentally measured development of the resistance between the two LM bodies (Fig. 4(b)), this process began with an insulating state between the two and subsequently went through a direct conducting state, which implied the existence of a period when the thickness of the liquid layer could reach $L_C$ and quantum tunneling condition could be satisfied. From a contrary way, such transient quantum tunneling conditions can also be expected from the cases of either satellite ejecting process or successive coalescence of the satellite droplets as well (Fig. 4(c)).

While the lubricating theory requires flowing insulating medium to generate a stable liquid layer, there exist another mechanism to prevent coalescing and to maintain a sandwich structure. This mechanism arises from the intrinsic electric-double-layer (EDL) structure of the liquid metal/electrolyte interface (Fig. 5). The liquid metal immersing in an electrolyte obtains a charged screen and two liquid metal droplets thus can be recognized as two like-charge bodies. Therefore, the droplets will experience repulsive force when they approach each other due to electrostatic interactions. The screening length $\lambda_D$ of the EDL can be obtained as [22]

$$\lambda_D = \sqrt{\frac{\varepsilon_0 \varepsilon_r N_A k_B T}{2F^2 I}} \in O(\text{nm}) \tag{6}$$

where $\varepsilon_0$ is the permittivity of free space, $\varepsilon_r$ the dielectric constant the electrolyte, $N_A$ the Avogadro constant, $k_B$ the Boltzmann constant, $T$ the temperature, $F$ the Faraday constant, and $I$ the ionic strength of the electrolyte. Since for most cases, $\lambda_D$ and $L_C$ are of the same scale, the electrostatic repulsion between the EDL of neighboring LM droplet will be significant around $L_C$ which may potentially help sustain the LM-IL-LM sandwich structure.

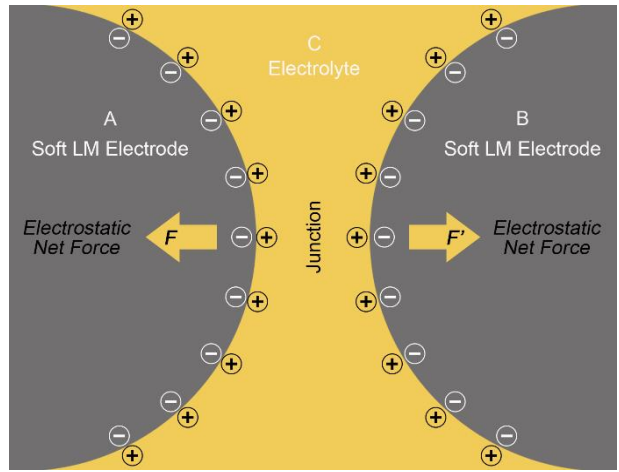

**Fig. 5.** Schematic drawing of the EDL structure of two adjacent liquid metal droplets in electrolyte and the repulsive force produced through electrostatic interaction of the charged surfaces.

### 3.3 All-soft quantum device with an insulating thin film

The insulating junction will not necessarily be liquid alone. That means, a soft layer can also be used as such junction for making transformable quantum device. In this sense, an all-soft quantum device (A-SQD) can be fabricated. In the above cases, oxidation of the LM surface is prevented by using dissolving electrolytes. As a matter of fact, the nonconductive oxide layer on the LM surface can serve as a naturally formed insulating layer. Importantly, the oxide layer formed on the LM under ambient atmosphere usually has a thickness around 1 nm [23] which is feasible for realizing quantum tunneling effect (Fig. 6(a)). Moreover, the thickness of the oxide layer can be adjusted through controlling temperature and oxygen content which means the tunneling junction can be flexibly tuned. The LM oxide film can also be preserved in liquid environment like oil, ethanol, surfactant solutions, etc. In Fig. 6(b) and (c), the large amount of closely packed LM droplets in a surfactant solution (Sodium dodecyl sulfate, SDS) has been experimentally demonstrated [14]. Consequently, the proposed soft LM quantum tunneling device can be made by directly combining two oxidized LM electrodes.

More generally, the junction can also be made of other types of thin films. For examples, thin insulating film of adjustable $L_C$ can be coated on the LM surface using physical or chemical vapor deposition methods. Also, self-assembled monolayers (SAMs) are typically within the $L_C$ scale and can be an alternative as well [23].

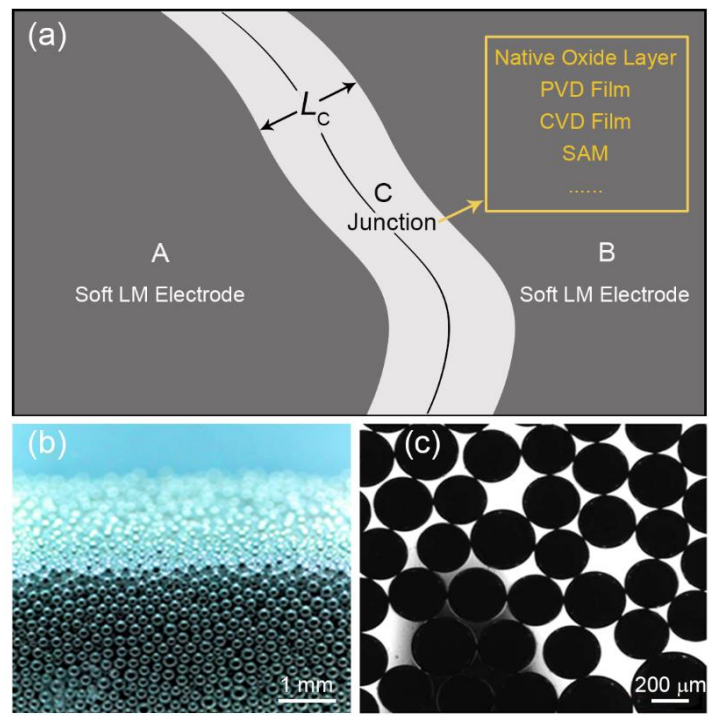

**Fig. 6.** (a) Schematics of all-soft LM quantum tunneling devices made by joining two liquid metal electrodes coated with thin insulating film; (b) Optical image of stable micro liquid metal droplets closely packed in SDS solution; (c) Microscopic image showing multiple point contact between micro LM droplets.

### 3.4 All-soft quantum device with a compressible insulating layer

Regarding the A-SQD, more technical strategies could help make such elements. Compressible materials such as elastomers and gels represent another class medium which can serve as the insulating layer for the proposed all-soft liquid metal quantum device. In what follows, we explain an easy going route to fabricate an array of transformable quantum tunneling devices within a LM droplet-hydrogel (Carbopol hydrogel) framework. As shown in Fig. 7(a), micrometer LM droplets can be well distributed in the hydrogel [15]. Micrometer distance between the outer surfaces of adjacent droplets ($d_0$) can be readily achieved (Fig. 7(b)) and with better control capability, i.e. sub micrometer distances are possible for realizing required functions. Since the hydrogel is capable of performing elastic deformation, the original distance $d_0$ can be reduced through compressing action. Using a liner elasticity assumption, the relation between the final distance $d$ and the external stress $\Delta p$ can be obtained as follows (Fig. 7(c)):

$$d = d_0(1 - \frac{\Delta p}{E}) \qquad (7)$$

where $E$ is the modulus of compression of the hydrogel. Fig. 7(d) plots $d$ against $\Delta p/E$ for different original distance $d_0$. For each curve, the adjustable range of $\Delta p/E$ is located in the filled region. It can be seen that small starting distance $d_0$ is favorable in order to compress the hydrogel layer to the quantum tunneling scale $L_C$ (0.1~10 nm, filled region in Fig. 7(d)). Also, a small $d_0$ is favorable for a wider adjustable range. Typically, $d_0$ should be of sub micrometer scale for practical thickness control. A rough estimate of the adjustable range of $\Delta p$ is on the scale of 10 to 100 Pa for hydrogels with $E$=100~1000 Pa.

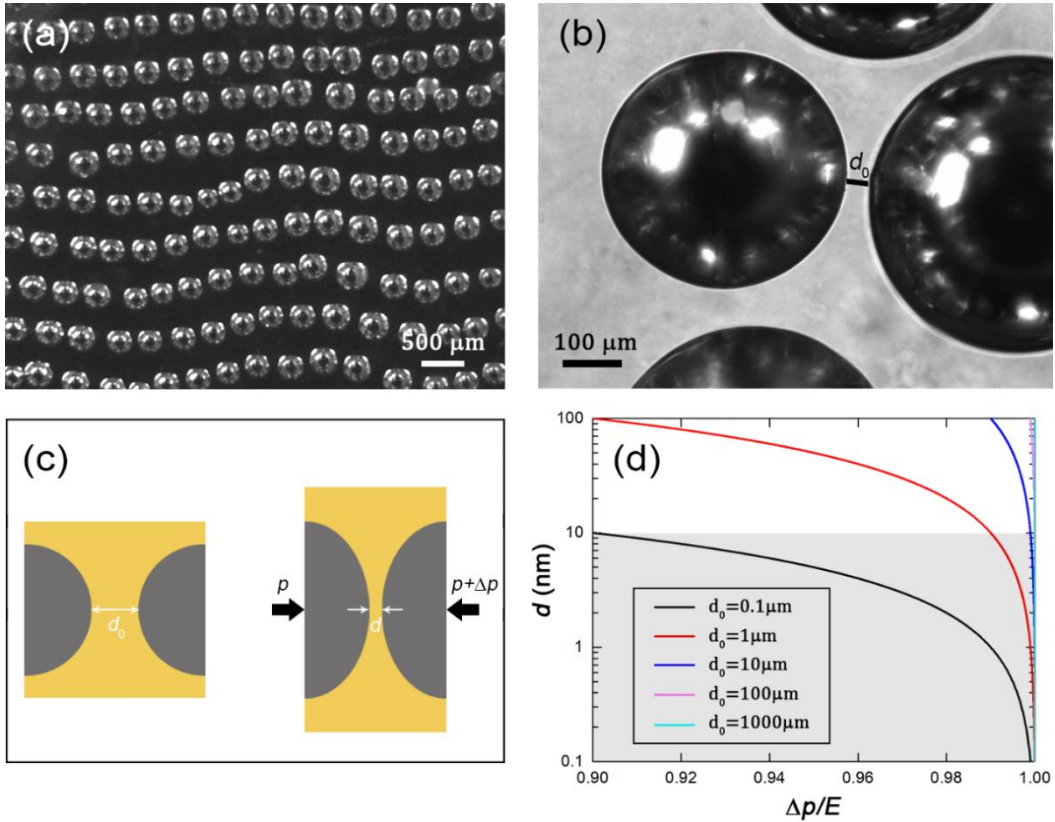

**Fig. 7.** (a) Microscope image of the LM extruded into Carbopol hydrogel [15]; (b) Microscale distance between LM droplets; (c) A schematic drawing shows exteral stress compresses the hydrogel layer; (d) The development of $d$ as a function of normalized stress $\Delta p/E$ for different original distance $d_0$. The filled gray rigion indicates $d$ enters the characteristic scale for quantum tunneling.

## 4. Discussion

In this work, strong evidence for quantum tunneling effect in the LM-IL-LM sandwich system is presented with potential quantum device configurations proposed, though there is no direct measurement of quantum tunneling effect in the current system. Straightforward advantages of the all-soft quantum devices over the traditional rigid ones are also illustrated. As summarized in Fig. 8, the insulating layer between LM electrodes can be water, oils, the native oxide layer of LM, SAMs, hydrogels and many other soft materials. Besides, the soft quantum tunneling devices will not be limited by the traditional sandwich structure and they can also work on various configurations. Since the LM can be transformed into various shapes, the LM electrodes can contact each other in different ways. For example, two spherical LM drops exhibit a point contact, two LM films exhibit plane contact, and two deformed LM drops may contact at more than one point. These manifold configuration modes can render more complex structure of all-soft quantum device.

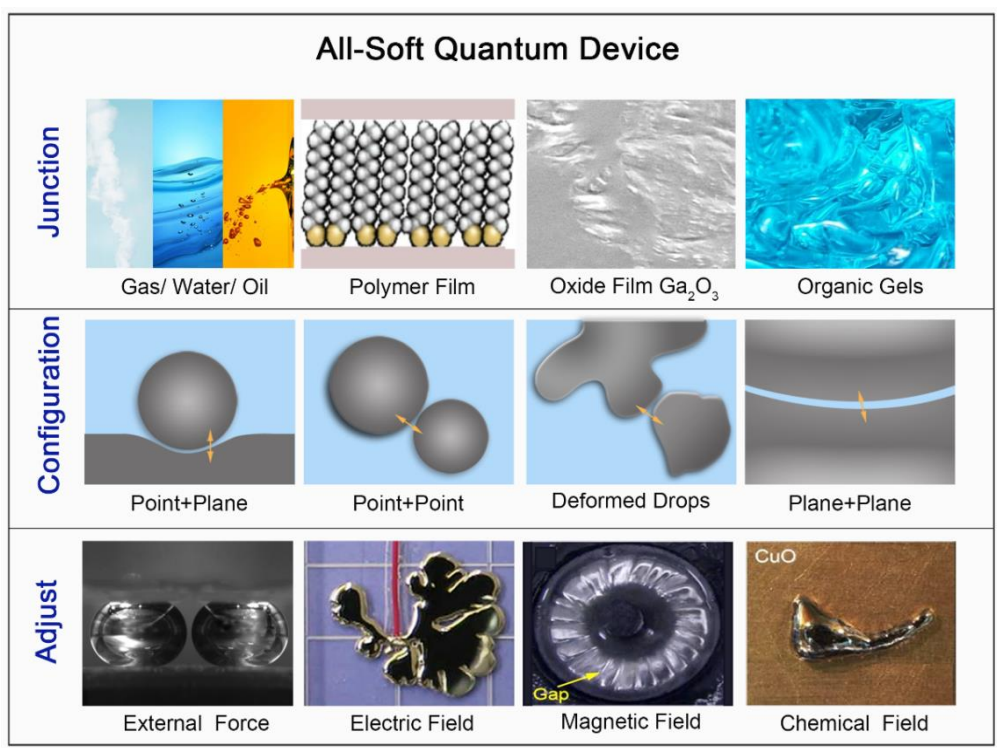

**Fig. 8.** All-soft quantum device: junction types, configurations and adjusting methods [18-20]

Overall, the present transformable all-soft quantum device owns a group of unique merits which are different from the traditional rigid-body one. And it implies important theoretical significance for developing future generation quantum devices. As outlined as below, for one particular case, the gallium-based room temperature LMs are quite suitable for manufacturing A-LQD in virtue of its unique advantages.

**High electric conductivity.** LMs generally own the inherent merit of good electric conductivity [24]. The electric conductivity of gallium-based LMs is on the order of $10^6$ S/m, which builds a basis of the LM as the electrode material of quantum devices.

**Highly smooth and uniform surface.** As we have mentioned in section 3.1, the quantum tunneling only occurs in the case of a barrier thickness of $0.1 - 10$ nm. For traditional solid electrodes of quantum devices, high quality of perfect surface flatness and uniformity are required, which significantly increases the processing difficulty and manufacturing cost. If the electrodes of quantum tunneling devices can be replaced by a material with highly smooth and uniform surface, these problems can be easily resolved. Studies have shown that LM has extremely smooth surface in the liquid environment compared to the solid surface, a hydrogen generation experiment demonstrates that bubbles are all produced from the liquid-solid interface while no bubble would appear at the LM-liquid interface since the latter provides little vaporization core [25]. Therefore, the LM-I-LM structure offers a naturally-smooth interface for making quantum tunneling junctions.

**Controllability and transformability.** In addition to utilize the external force to directly change the droplet space, numerous external fields, including electric field, magnetic field, chemical field, optical field, etc. can all be possibly adopted to manipulate the droplet shape and the structure of the LM electrodes, thus the whole A-LQD is highly flexible, changeable and tunable. This easy manipulation character greatly extends the diversification and compatibility of A-LQD. In addition, since the electrolyte interface can be tuned into rather small nanometer size while its surrounding two neighboring liquid metal electrodes can be very large. In this sense, the current A-LQD offers a straight forward to make macroscale quantum device which will be rather easy to handle and operate.

**Flexibility and diversity.** The LM-I-LM sandwich structure is not just a changeless framework. It can in fact be an adaptive combination of a variety of transformable structures. For example, a series of quantum tunneling devices can be formed spontaneously when LM droplets with different sizes are immersed in a surfactant solution or a supporting hydrogel (Fig. 7(a,b)), because the LM droplets are separated by a thin layer of insulating liquid [14, 15]. Additionally, the droplet size, droplet space and solution type are all adjustable. These features present the LM quantum device an unusually high degree of freedom, so as to provide a convenient and efficient way to the fabrication of quantum devices with complex structures. In the coming time, through constructing specific droplet arrays with designed logistic structures, the current A-LQD or A-SQD can be developed as intelligent quantum device including new quantum computational systems.

**Easy manipulation:** A significant feature of LM still lies in that its surface tension is pretty high, which is almost ten times that of water (~700 mN/m) [26]. Consequently, a surface tension gradient across the LM body will result in a propelling force, which is sufficient to cause droplet deformation and even drive small droplets. Some external fields such as electrical field [27], magnetic field [28], chemical field [11, 29], optical field [30], electromagnetic field [31] and combined fields [32] etc. have all been found to effectively affect the deformation and movement of LM through intentionally altering its surface tension. And the controllability is pretty strong. For example, in certain case, through applying an electric field, a large LM film can quickly transform into a tiny sphere, with over one thousand times variation in specific surface area [27]. Accompanied by the chemical field, LM can perform even more complex transformations. It is found that LM droplet changes to a flat and dull puddle when deposited on a graphite surface, which is easier to be manipulated as various stable shapes with sharp angles [12]. With these means to flexibly manipulate LM objects in LM-based quantum devices, the shape of the LM electrodes and the liquid interlayer thickness can be adjusted and regulated to obtain the desired shape and function.

## 5. Conclusion

The all soft or liquid quantum device is a fundamentally conceptual innovation over traditional rigid quantum devices, metal materials and understanding of solution characteristics. Such type of unconventional quantum device owns a group of unique advantages such as high flexibility, large deformability, convenient transformability and easy manufacturability, including straightforward manipulation. This is highly conducive to the polarization of low-cost quantum device and may promote the development of future generation quantum technology. Further studies should focus on the way to accurately control the thickness of the liquid interlayer, regulate the whole A-LQD or A-SQD structure, and achieve a group of fully controllable transformable quantum devices including developing their applications in intelligent and quantum computing systems.

## Acknowledgment

This work is partially supported by the Dean's Research Funding and the Frontier Project of the Chinese Academy of Sciences.

## References

1. Dirac PAM. The Principles of Quantum Mechanics [M]. Oxford University Press, 1981.
2. Serway R, Faughn J, Vuille C. College Physics [M]. Cengage Learning, 2008.
3. Trixler F. Quantum tunneling to the origin and evolution of life [J]. Current Organic Chemistry, 2013, 17(16): 1758-1770.
4. Taylor J R, Dubson M A, Zafiratos C D. Modern Physics for Scientists and Engineers [M]. Prentice-Hall, 2004.
5. Razavy M. Quantum Theory of Tunneling [M]. World Scientific, 2013.
6. Lerner R G, Trigg G L, Rigden J S. Encyclopedia of Physics [M]. Wiley-VCH, 1991.
7. Zhao X, Xu S, Liu J. Surface tension of liquid metal: role, mechanism and application [J]. Frontiers in Energy, 2017. DOI: 10.1007/s11708-017-0463-9
8. Tang J, Zhao X, Li J, et al. Liquid metal phagocytosis: Intermetallic wetting induced particle internalization [J]. Advanced Science, 2017, 4(5): 1700024.
9. Tang J, Wang J, Liu J, et al. Jumping liquid metal droplet in electrolyte triggered

by solid metal particles [J]. Applied Physics Letters, 2016, 108: 223901.
10. Zhang J, Yao Y, Sheng L, et al. Self‐fueled biomimetic liquid metal mollusk [J]. Advanced Materials, 2015, 27(16): 2648-2655.
11. Zavabeti A, Daeneke T, Chrimes A F, et al. Ionic imbalance induced self-propulsion of liquid metals. Nature Communications, 2016, 7: 12402.
12. Hu L, Wang L, Ding Y, et al. Manipulation of liquid metals on a graphite surface [J]. Advanced Materials, 2016, 28(41): 9210-9217.
13. Yi L, Ding Y, Yuan B, et al. Breathing to harvest energy as a mechanism towards making a liquid metal beating heart [J]. RSC Advances, 2016, 6: 94692.
14. Yu Y, Wang Q, Yi L, et al. Channelless fabrication for large-scale preparation of room temperature liquid metal droplets [J]. Advanced Engineering Materials, 2014, 16(2): 255-262.
15. Yu Y, Liu F, Zhang R, et al. Suspension 3D printing of liquid metal into self‐healing hydrogel [J]. Advanced Materials Technologies, 2017: 1700173. https://doi.org/10.1002/admt.201700173
16. Sen D. The uncertainty relations in quantum mechanics [J]. Current Science, 2014, 107(2): 203-218.
17. Heisenberg W. The Physical Principles of the Quantum Theory [M]. Courier Corporation, 1949.
18. Greiner W. Quantum Mechanics [J]. Springer Berlin, 1994, 71(1): 1–24.
19. Eisberg R, Resnick R, Brown J. Quantum physics of atoms, molecules, solids, nuclei, and particles [J]. Physics Today, 1986, 39: 110.
20. Zhao X, Tang J, Liu J. Surfing liquid metal droplet on the same metal bath via electrolyte interface [J]. Applied Physics Letters, 2017, 111(10): 101603.
21. Couder Y, Fort E, Gautier C H, et al. From bouncing to floating: Noncoalescence of drops on a fluid bath. Physical Review Letters, 2005, 94(17): 177801.
22. Devanathan M A V, Tilak B. V. K. S. R. A. The structure of the electrical double layer at the metal-solution interface. Chemical Reviews, 1965, 65(6): 635-684.
23. Thuo M M, Reus W F, Nijhuis C A, et al. Odd-even effects in charge transport across self assembled monolayer based Metal-SAM-Metal junctions. Journal of the American Chemical Society, 2011, 133: 2962-2975.
24. Sen P, Kim C-J C. Microscale liquid-metal switches—A review [J]. IEEE Transactions on Industrial Electronics, 2009, 56(4): 1314-1330.
25. Yuan B, Tan S, Liu J. Dynamic hydrogen generation phenomenon of aluminum fed liquid phase Ga–In alloy inside NaOH electrolyte [J]. International Journal of Hydrogen Energy, 2016, 41(3): 1453-1459.
26. Zrnic D, Swatik D. On the resistivity and surface tension of the eutectic alloy of gallium and indium [J]. Journal of the Less Common Metals, 1969, 18(1): 67-68.
27. Sheng L, Zhang J, Liu J. Diverse transformations of liquid metals between different morphologies [J]. Advanced Materials, 2014, 26(34): 6036-6042.
28. Tan S-C, Gui H, Yuan B, et al. Magnetic trap effect to restrict motion of self-powered tiny liquid metal motors [J]. Applied Physics Letters, 2015, 107(7): 071904.
29. Zhang J, Sheng L, Liu J. Synthetically chemical-electrical mechanism for

controlling large scale reversible deformation of liquid metal objects [J]. Scientific Reports, 2014, 4: 7116.

30. Tang X, Tang S-Y, Sivan V, et al. Photochemically induced motion of liquid metal marbles [J]. Applied Physics Letters, 2013, 103(17): 174104.
31. Wang L, Liu J. Liquid metal folding patterns induced by electric capillary force [J]. Applied Physics Letters, 2016, 108(16): 161602.
32. Hu L, Li J, Tang J, et al. Surface effects of liquid metal amoeba [J]. Science Bulletin, 2017(10): 700-706.